\documentclass[aps,floatfix,superscriptaddress,showpacs]{revtex4}

\usepackage{epsfig}
\usepackage{amsmath}
\usepackage{graphicx,psfrag}
\usepackage{dcolumn}
\usepackage{bm}
\usepackage{slashed}
\usepackage{color}

\newcommand{\beq}{\begin{equation}}
\newcommand{\eeq}{\end{equation}}
\newcommand{\bea}{\begin{eqnarray}}
\newcommand{\eea}{\end{eqnarray}}
\newcommand{\hf} {\frac{1}{2}}

\newcommand{\nn}{\nonumber\\}

\newcommand\fig[1]     {Fig.\,{\ref{#1}}}

\newcommand\app[1]     {Appendix~\ref{#1}}

\def\eq#1{(\ref{#1})}
\def\s0#1#2{\mbox{\small{$ \frac{#1}{#2} $}}}
\def\0#1#2{\frac{#1}{#2}}

\def\mr#1{{\mathrm{#1}}}

\begin{document}

\title{Efficiency of magnetic hyperthermia in the presence of rotating and static fields} 

\author{Zs. Isz\'aly} 
\affiliation{University of Debrecen, H-4010 Debrecen P.O.Box 105, Hungary}

\author{K. Lov\'asz}
\affiliation{University of Debrecen, H-4010 Debrecen P.O.Box 105, Hungary}

\author{I. Nagy}
\affiliation{University of Debrecen, H-4010 Debrecen P.O.Box 105, Hungary}

\author{I. G. M\'ari\'an}
\affiliation{University of Debrecen, H-4010 Debrecen P.O.Box 105, Hungary}

\author{J. R\'acz}
\affiliation{University of Debrecen, H-4010 Debrecen P.O.Box 105, Hungary}

\author{I. A. Szab\'o}
\affiliation{University of Debrecen, H-4010 Debrecen P.O.Box 105, Hungary}

\author{L. T\'oth}
\affiliation{University of Debrecen, H-4010 Debrecen P.O.Box 105, Hungary}

\author{N. F. Vas}
\affiliation{University of Debrecen, H-4010 Debrecen P.O.Box 105, Hungary}

\author{V. V\'ekony}
\affiliation{University of Debrecen, H-4010 Debrecen P.O.Box 105, Hungary}

\author{I. N\'andori}
\affiliation{University of Debrecen, H-4010 Debrecen P.O.Box 105, Hungary}
\affiliation{MTA-DE Particle Physics Research Group, P.O.Box 51, H-4001 Debrecen, Hungary}
\affiliation{MTA Atomki, P.O. Box 51, H-4001 Debrecen, Hungary}

\begin{abstract} 
Single-domain ferromagnetic nanoparticle systems can be used to transfer energy from a time-dependent 
magnetic field into their environment. This local heat generation, i.e., magnetic hyperthermia, receives
applications in cancer therapy which requires the enhancement of the energy loss. A possible way to improve the 
efficiency is to chose a proper type of applied field, e.g., a rotating instead of an oscillating one. The latter case is 
very well studied and there is an increasing interest in the literature to investigate the former although it is still unclear 
under which circumstances the rotating applied field can be more favourable than the oscillating one. The goal of this
work is to incorporate the presence of a static field and to perform a systematic study of the non-linear dynamics 
of the magnetisation in the framework of the deterministic Landau-Lifshitz-Gilbert equation in order to calculate 
energy losses. Two cases are considered: the static field is either assumed to be perpendicular to the plane of 
rotation or situated in the plane of rotation. In the latter case a significant increase in the energy loss/cycle is 
observed if the magnitudes of the static and the rotating fields have a certain ratio (e.g. it should be one for isotropic 
nanoparticles). It can be used to "super-localise" the heat transfer: in case of an inhomogeneous applied static field, 
tissues are heated up only where the magnitudes of the static and rotating fields reach the required ratio.
\end{abstract}

\maketitle

\section{Introduction}
\label{sec_intro}
Relaxation of ferromagnetic nanoparticle systems under alternating external applied fields is a very active 
research field, both in its theoretical and material-science aspects. Recently, a systematic study of magnetic nanoparticle 
response to a time-dependent applied field and the corresponding energy transfer have been performed in Ref.\cite{lyutyy_general} 
taking into account the interplay between the mechanical motion of nanoparticles with respect to their environment and the internal 
dynamics of their magnetisation vector with respect to the crystal lattice. Indeed, two different approximations can be used to study 
and maximise the energy loss, namely the "rigid dipole" when the magnetisation vector is assumed to be rigidly attached to the 
crystal and only the particle can move or rotate and the "fixed nanoparticle" approach where the nanoparticles are immobilised and 
only the internal motion of the magnetisation vector is allowed. It is not obvious to describe the joint dynamics of a nanoparticle by 
the combination of these two types of motion. However, after some attempts \cite{joined_motion} a set of equations of motion was 
proposed \cite{joined_motion_2}. Furthermore, finite anisotropy effects are also investigated in \cite{lyutyy_general} for the case 
of periodically driven ferromagnetic nanoparticles and even thermal effects and dipole interactions are taken into account in 
\cite{thermal_and_dipole}. All these considerations are important steps towards an improved magnetic hyperthermia by chosing 
a proper type of applied field, i.e.,  either the rotating or the oscillating one. 

There is, however, an important issue namely the presence of a static applied field which has not been involved yet in a study of 
power loss of periodically driven magnetic nanoparticles, although it is easy to realise in practice and, as we show in this work, it 
has important impact. Therefore, in this work we consider and maximise the efficiency of magnetic hyperthermia in the presence 
of rotating and static fields. We restrict our consideration to the case of fixed particles where the deterministic Landau-Lifshitz-Gilbert 
equation is used to calculate the energy loss. Thus, we are using two important approximations the "fixed particle" approach and 
the deterministic (and not stochastic) description. The former is validated by Ref \cite{lyutyy_general} where it was shown that at least 
for small oscillations the damping precession of the nanoparticle magnetic moment is the main source of energy dissipation,
however, the frequency range used for hyperthermia do not allow us to neglect the mechanical rotation of the 
nanoparticles if the particles size is small ($\sim 20$ nm) which means that our fixed particle approach for hyperthermia applications 
is insufficient for very small nanoparticles. The latter constraint is related to the fact that no thermal fluctuations are taken into 
account in this work which represents a reliable approximation, for circularly polarised applied field due to the presence of steady 
state motions. Indeed it was shown in \cite{Raikher} for isotropic and fixed nanoparticles that the power losses of the rotating applied
field with and without a stochastic description are very close to each other, in other words, stochastic description is less important in 
this case. In addition, it was also demonstrated that the heating efficiency of the rotating and oscillating applied fields are almost 
identical in the limit of very small frequency required for hyperthermia \cite{Ahsen2010, Raikher}.

Let us also note, that magnetic hyperthermia or more general, the nonlinear dynamics of the magnetization in single-domain 
ferromagnetic nanoparticles receives increasing interest in various fields including data storage, biomedical engineering, spintronics,  
\cite{berger,scherer,feldhof,muller,leschorn,schlio1,schlio2,rosensweig,schlio3,coffey,pankhurst,hergt1,hergt2,vallejo,biomedical}, 
and as mentioned in the abstract in cancer therapy, \cite{Stauffer, Johannsen,Bellizzi}. A possible way to improve the efficiency of magnetic 
hyperthermia is to chose a proper type of applied field. For example, in \cite{heat_enhance} a special rotating field has been 
proposed where the direction of rotation is changed in every cycle. The goal of this work is to study the effect of a static 
magnetic field on the heat generation by isotropic and anisotropic nanoparticles driven by a rotating field. Two different 
orientations are considered. In the first case, the static field is assumed to be perpendicular to the plane of rotation and the 
special situation where the direction of rotation is changed in every cycle is also considered. In the second case, the static field 
is situated in the plane of rotation.

The paper is organised as follows. After the introduction, in Sec.~\ref{sec_llg_zero_stat} the deterministic Landau-Lifshitz-Gilbert 
(LLG) equation has been introduced and the previous results obtained in the absence of static field is summarised very briefly. 
The study of the role of a non-zero static field is the main goal of this work, and the findings are discussed in two sections. The 
static field is assumed to be (i) perpendicular to the plane of rotation Sec.~\ref{sec_nonzero_stat}, (ii) situated in the plane of 
rotation Sec.~\ref{sec_nonzero_stat_in_plane}. Finally, Sec.~\ref{sec_sum} stands for the conclusions.

\section{The LLG equation and previous results}
\label{sec_llg_zero_stat}
In this work we use the deterministic Landau-Lifshitz-Gilbert (LLG) equation \cite{LL,Gilbert} which is suitable to describe the
dynamics in case of a fixed particle approach. It retains the magnitude of the magnetisation, so introducing a unit vector 
${\bf M} = {\bf m}/m_S$, $m_S$ with the saturation magnetic moment the LLG equation reads
\begin{equation}
\label{LLG}
\frac{\rm{d}}{\rm{d}t} {\bf M} = -\gamma' [{\bf M \times H_{\rm{eff}}}] 
+ \alpha' [[{\bf M\times H_{\rm{eff}}]\times M}],
\end{equation}
with $\gamma' = \mu_0 \gamma_0 /(1+\alpha^2)$, $\alpha' = \gamma' \alpha$ and the dimensionless 
damping $\alpha = \mu_0\gamma_0\eta m_S$ (e.g. $\alpha = 0.1$ and $\alpha = 0.3$ have been used in 
\cite{Lyutyy_energy} and \cite{Giordano}, respectively). The effective field is defined as 
\bea
\label{H}
{\bf H}_{\rm{eff}} = H_0 \, \, (\cos(\omega t), \sin(\omega t), \lambda_{\rm{eff}} M_z + b_0),
\eea
where $\lambda_{\rm{eff}} = H_a /H_0$ is the anisotropy ratio and $H_0 b_0$ stands for the static field. In some figures, 
we use the notation $H \equiv H_0$. In this work we use a uniaxial anisotropy. If $\lambda_{\rm{eff}} > 0$ 
($H_a > 0$), the anisotropy will turn the magnetisation towards the z-axis, if  $\lambda_{\rm{eff}} < 0$ ($H_a < 0$), into 
the $xy$-plane.  Furthermore, it is straightforward to rewrite the LLG equation in polar coordinates ($\theta,\varphi$) in 
the reference frame attached to the rotating field by introducing  $\omega_L = H_0 \gamma'$ and $\alpha_N = H_0 \alpha'$. 
A good choice for parameters suitable for hyperthermia (with $\alpha = 0.1$ and $H_0 = 18$ kA/m), are 
$\omega = 5 \times 10^5 \, \mr{Hz}$, $\omega_L = 4 \times10^9 \, \mr{Hz}$ and $\alpha_N =  4 \times 10^8 \, \mr{Hz}$ 
where $\lambda_{\mr{eff}}$ is the dimensionless anisotropy parameter.  By introducing a dimensionless "time" 
$\tilde t = t/t_0$ with $t_0 = 0.5 \times 10^{-10}$s one can introduce  the dimensionless parameters such as
$\omega \to  \omega t_0 = 2.5 \times 10^{-5}$, $\omega_L \to \omega_L t_0  = 0.2$ and $\alpha_N \to \alpha_N t_0  = 0.02$. 
It is also useful to rewrite the LLG equation for Descartes coordinates of the magnetisation vector ($u_x, u_y, u_z$) in the
rotating frame which reads as
\bea
\label{rot_u}
\frac{du_{x}}{dt} &=& \omega u_{y}-u_{y}\omega_{L}b_{0} + \alpha_{N}u_{y}^2 + \alpha_{N}u_{z}(u_{z} - b_{0}u_{x}) \nn
\frac{du_{y}}{dt} &=& -\omega u_{x} -\omega_{L}u_{z} + \omega_{L}u_{x}b_{0} - \alpha_{N}u_{x}u_{y} - \alpha_{N}u_{z}u_{y}b_{0} \nn
\frac{du_{z}}{dt} &=& \omega_{L}u_{y} - \alpha_{N}u_{x}u_{z} + \alpha_{N}b_{0}(1-u_{z}^2). 
\eea
which are not independent since Eq.~\eq{LLG} stands for a unit vector. 

Let us first summarise very briefly the main results of previous investigations on rotating applied field in the absence
of any static field \cite{aniso,heat_enhance}, i.e., for $b_0 = 0$. We recall results on (i) the steady state solution of the 
LLG equation which can also be understood as the attractive fixed point solution of the LLG equation in the rotating frame,
(ii) out of the steady state solutions which are achieved by changing the direction of the rotation in every full cycle.

\subsection{Steady state solution in the absence of static fields}
\begin{figure}[ht] 
\begin{center} 
\includegraphics[width=6cm]{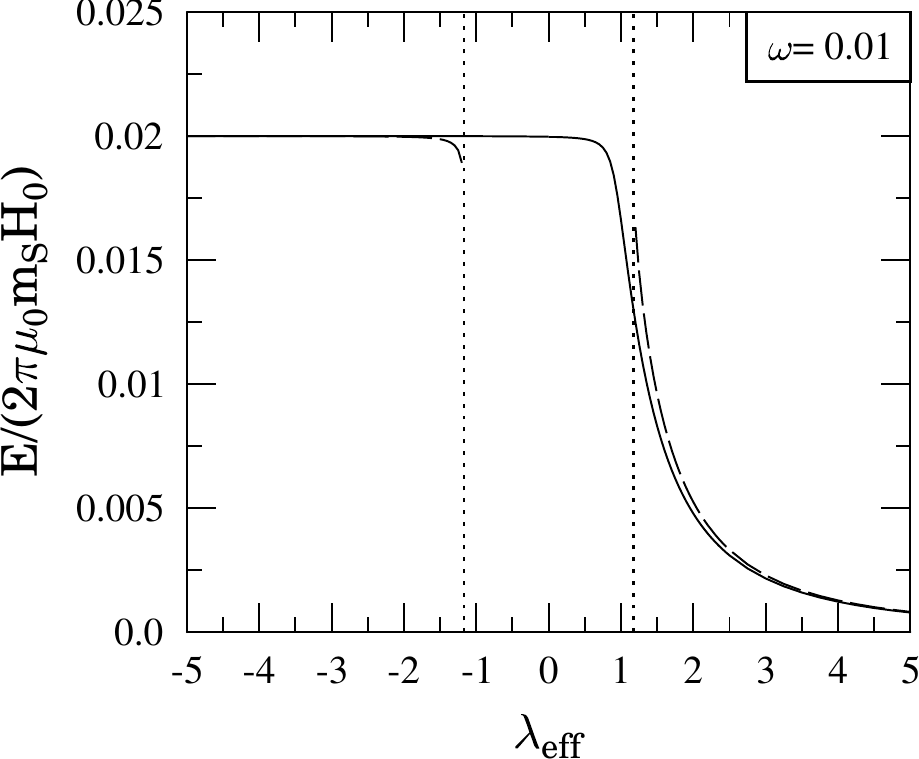}
\caption{The energy loss per cycle as a function of the anisotropy parameter $\lambda_{\mr{eff}}$ is shown for $\alpha_N = 0.1$, 
$\omega_L = 0.2$. Vertical lines indicate the critical value $\lambda_c$. Solid and dashed lines correspond to energy losses of 
steady state solutions. 
\label{fig1}
} 
\end{center}
\end{figure}

The steady state solution of the LLG equation in the absence of any static field in the inertial frame has the following form
\begin{eqnarray}
\label{steady_state}
M_{x}(t) &=& 
u_{x0} \cos(\omega t) - u_{y0} \sin(\omega t),
\nonumber \\
M_{y}(t) &=&
 u_{x0} \sin(\omega t) + u_{y0} \cos(\omega t),
\nonumber \\
M_{z}(t) &=& u_{z0}.
\end{eqnarray}
where $u_{x0}$ and $u_{y0}$ are functions of the parameters $\omega$, $\omega_L$, $\alpha_N$ and $\lambda_{\rm{eff}}$
determined the attractive fixed point solution of the LLG equation for $b_0 = 0$. Below a critical value of the anisotropy 
$\vert\lambda_{\mr{eff}}\vert < \lambda_c$, only a single attractive fixed point appears. Above this critical anisotropy one
finds two steady state solutions. The energy loss is calculated using these attractive fixed point solutions in the formula for
the energy dissipated in a single cycle
\beq
\label{def_loss}
E = \mu_0 m_S \int_{0}^{\frac{2\pi}{\omega}} \mr{d}t 
\left({\bf H}_{\rm{eff}} \cdot \frac{d{\bf M}}{dt} \right)
= \mu_0 2\pi m_S H_0 (-u_{y0}),
\nonumber
\eeq
which can be taken in the low-frequency $\omega \ll \alpha_N$ and small anisotropy $\vert \lambda_{\mr{eff}} \vert \ll~1$ limits
where $\lambda_{\mr{eff}}$ appears only at the next-to-leading terms. This is demonstrated in \fig{fig1} which shows that for 
$\vert\lambda_{\mr{eff}}\vert < \lambda_c$ the energy loss per cycle becomes identical to that in the isotropic case, independently 
of the sign of $\lambda_{\rm{eff}}$. 
%
%
For isotropic nanoparticles the oscillating field was found to be more favourable than the rotating one \cite{Chatel,Raikher}
and the anisotropy cannot change this picture. Therefore, one has to try to move out of the steady state solutions.

\subsection{Out of the steady state solution in the absence of static fields}
Let us calculate the energy loss out of the steady states. The reason for doing this is related to Ref. \cite{heat_enhance},
where it was shown that steady state motions always correspond to minimal dissipation. In other words, enhancement of heat generation 
is possible if the motion of magnetic moment is out of steady states which are related to attractive fixed points in the rotating frame, thus, 
the magnetic moment tends to a steady state very rapidly (independently of its initial conditions). Out of steady state motion can be achieved 
by an appropriate applied field when, for example, the direction of rotation is changed periodically. In order to demonstrate that steady state
motions correspond to minimal dissipation one has to consider energy loss obtained in the very first cycle.

\begin{figure}[ht] 
\begin{center} 
\includegraphics[width=7cm]{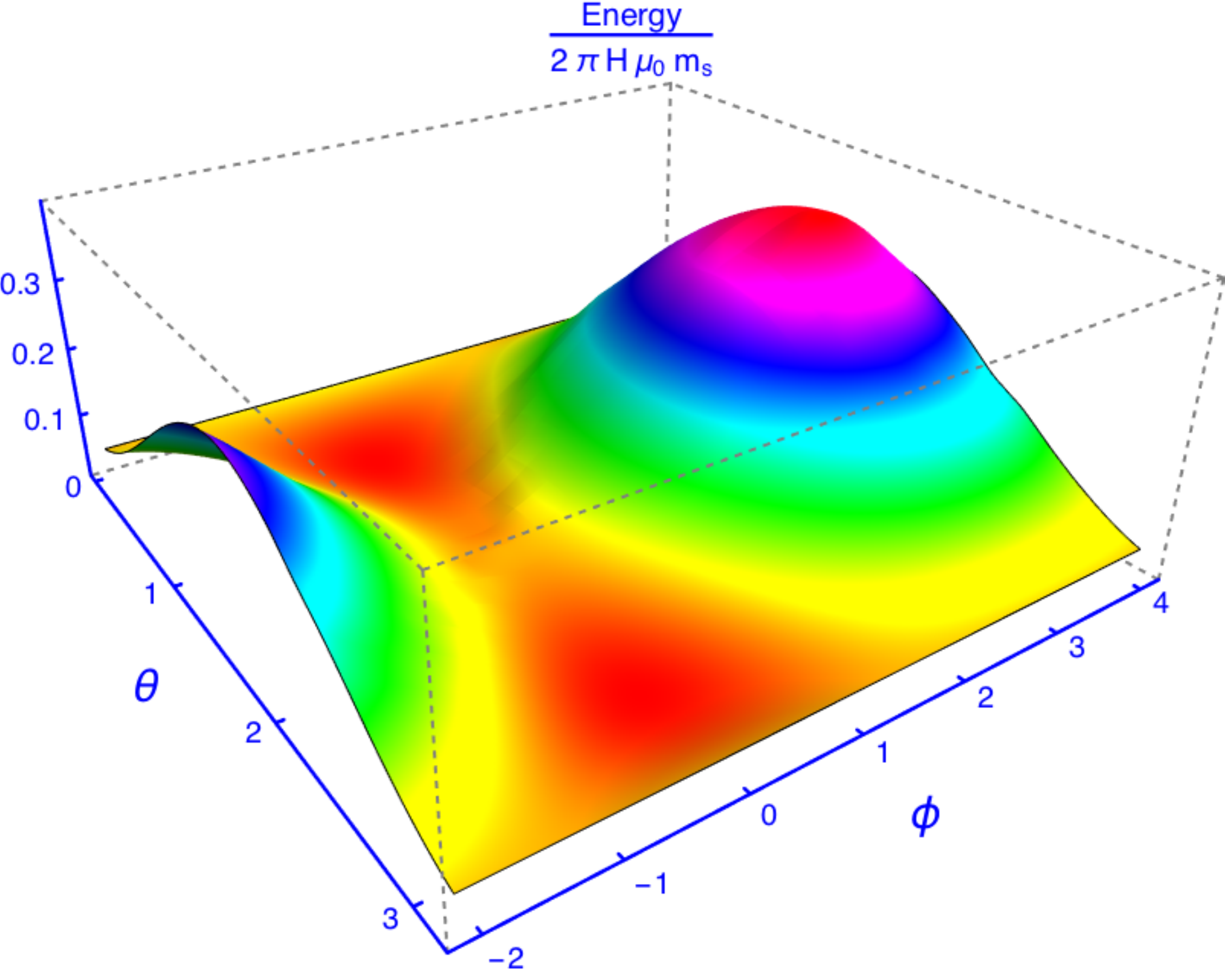}
\caption{(Color online). Energy loss obtained in the first cycle, out of the steady state for $b_0=0$ as a function of initial 
conditions on the ($\theta,\phi$) plane with $\lambda_{\mr{eff}} = 1.9$, $\alpha_N = 0.1$, $\omega_L = 0.2$ and $\omega = 0.01$.
\label{fig2}
} 
\end{center}
\end{figure}

The 3D-plot of \fig{fig2} shows the energy loss  obtained in the first cycle of the external field as a function of various starting points on 
the ($\theta,\phi$) plane. 
%
%
In \fig{fig2} the "wells" correspond to attractive fixed points which produce us the lowest loss energy per cycle while the "hill" 
with the maximal energy loss is related to initial conditions taken at the repulsive fixed point. This indicates that it is worth to 
move out of the steady state which corresponds to minimal dissipation, i.e., wells of \fig{fig2}. 

\begin{figure}[ht] 
\begin{center} 
\includegraphics[width=6cm]{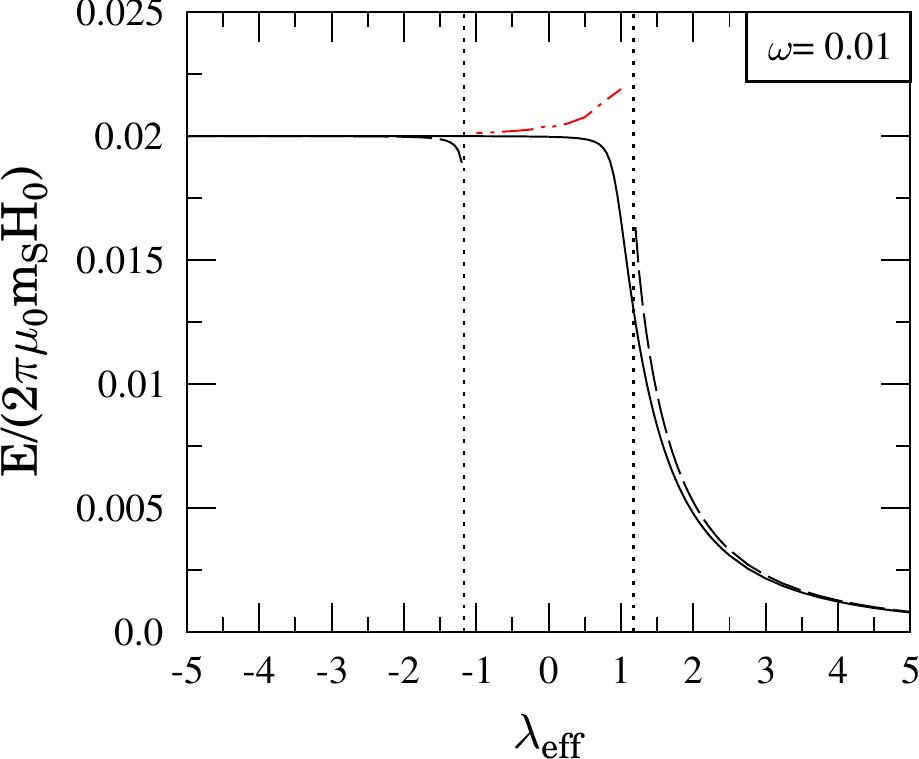}
\caption{(Color online). The same as \fig{fig1} but extended by a solution (red dahsed-dotted line) for a rotating field with 
alternating direction.  
\label{fig3}
} 
\end{center}
\end{figure}

The idea is to use a rotating field with alternating direction, i.e., in every full cycle the angular velocity in \eq{H} changes its
sign ($\omega \to -\omega$). In this way the magnetic moment is removed from its steady state in each cycle and the energy
loss is found to be larger than that of the steady state, see \fig{fig3}.
%
%

The increase of the energy loss near the critical anisotropy ($\lambda_{\mr{eff}} \sim \lambda_c$) can reach a value that is 
about 15\% or 100\% larger as compared to the isotropic case in the limit of low frequency ($\omega = 0.001$) and lower 
values of the damping, i.e. $\alpha_N = 0.05$ or $\alpha_N = 0.01$, see Fig.4 of \cite{heat_enhance}.

\section{Static field perpendicular to the plane of rotation}
\label{sec_nonzero_stat}
New findings of the this work are presented in this section. Our goal is to clarify the role of the stabilising static field
on the possible enhancement of energy loss. The static field is assumed to perpendicular to the plane of rotation, 
i.e., the effective applied field is defined by Eq.~\eq{H} with $b_0 \neq 0$. Similarly to the previous section, two different
regimes are studied: energy losses (i) in case of steady states, (ii) out of steady states.

\subsection{Steady state solution}
\begin{figure}[ht] 
\begin{center} 
\includegraphics[width=7.5cm]{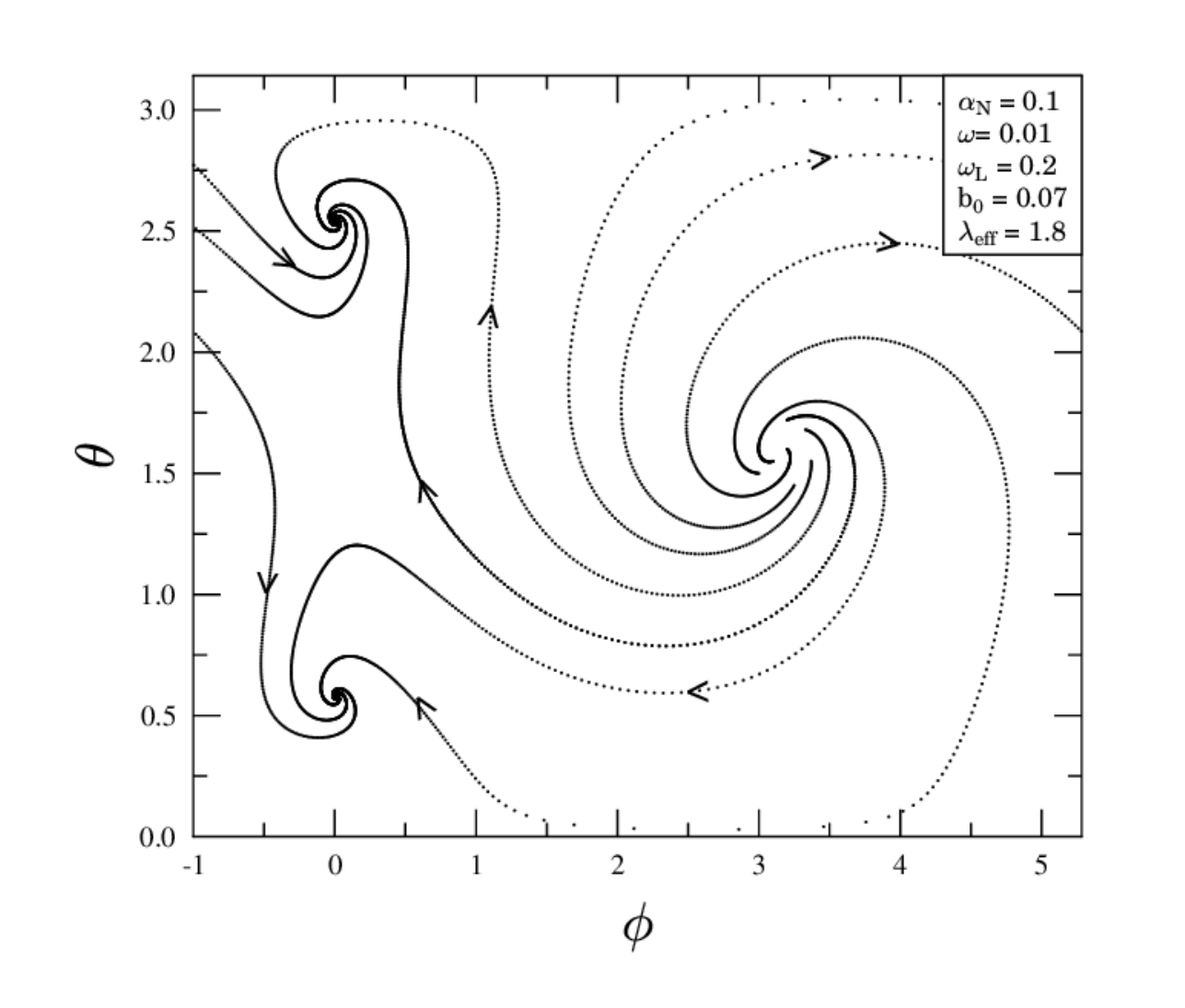}
\caption{Orbit map in the rotating frame for $b_0 \neq 0$.
\label{fig4}
} 
\end{center}
\end{figure}

Let us first solve the LLG equation numerically which immediately indicates the existence of steady state solutions
for $b_0 \neq 0$ which correspond to attractive fixed points of the orbit map \fig{fig4} obtained in the rotating frame.
%
%

The steady state solutions can be determined by the fixed point equation of \eq{rot_u} where the derivatives set to
zero and the resulting set of algebraic equations can be written in a single equation for $u_z$ which reads
\bea
u^2_{z} = (1-u^2_{z})  &\Bigl[\left(-\frac{\omega_{L}\omega}{\omega^2_{L} + \alpha^2_{N}} + b_{0} + \lambda_{eff}u_{z}\right)^2 
+ \left(\frac{\alpha_{N}\omega}{\omega^2_{L} + \alpha^2_{N}} u_{z}\right)^2\Bigr]. 
\eea
It can be solved analytically and the energy loss can be given which has the following form in the low-frequency 
$\omega \ll \alpha_N$ limit for vanishing anisotropy $\lambda_{\mr{eff}} = 0$, 
\bea
\label{loss_low-freq_static}
E \approx 2 \pi \mu_0 m_S H_0 \Bigl[\frac{\alpha_N \omega}{(\omega_L^2 + \alpha_N^2) (1 + b_0^2)}
+\frac{2\alpha_N  \omega_L b_0 \omega^2}{(\omega_L^2 + \alpha_N^2)^2 (1 + b_0)^2} \Bigr].
\eea
There is an important difference between the dependence of the energy loss on the strength of the static field $b_0$ in \eq{loss_low-freq_static},
and on the anisotropy ratio $\lambda_{\mr{eff}}$. Namely, the former appears in the leading order term 
in the energy loss formula, thus, it does not vanish for small frequencies, i.e., in the limit suitable for hyperthermia. 

\begin{figure}[ht] 
\begin{center} 
\includegraphics[width=7cm]{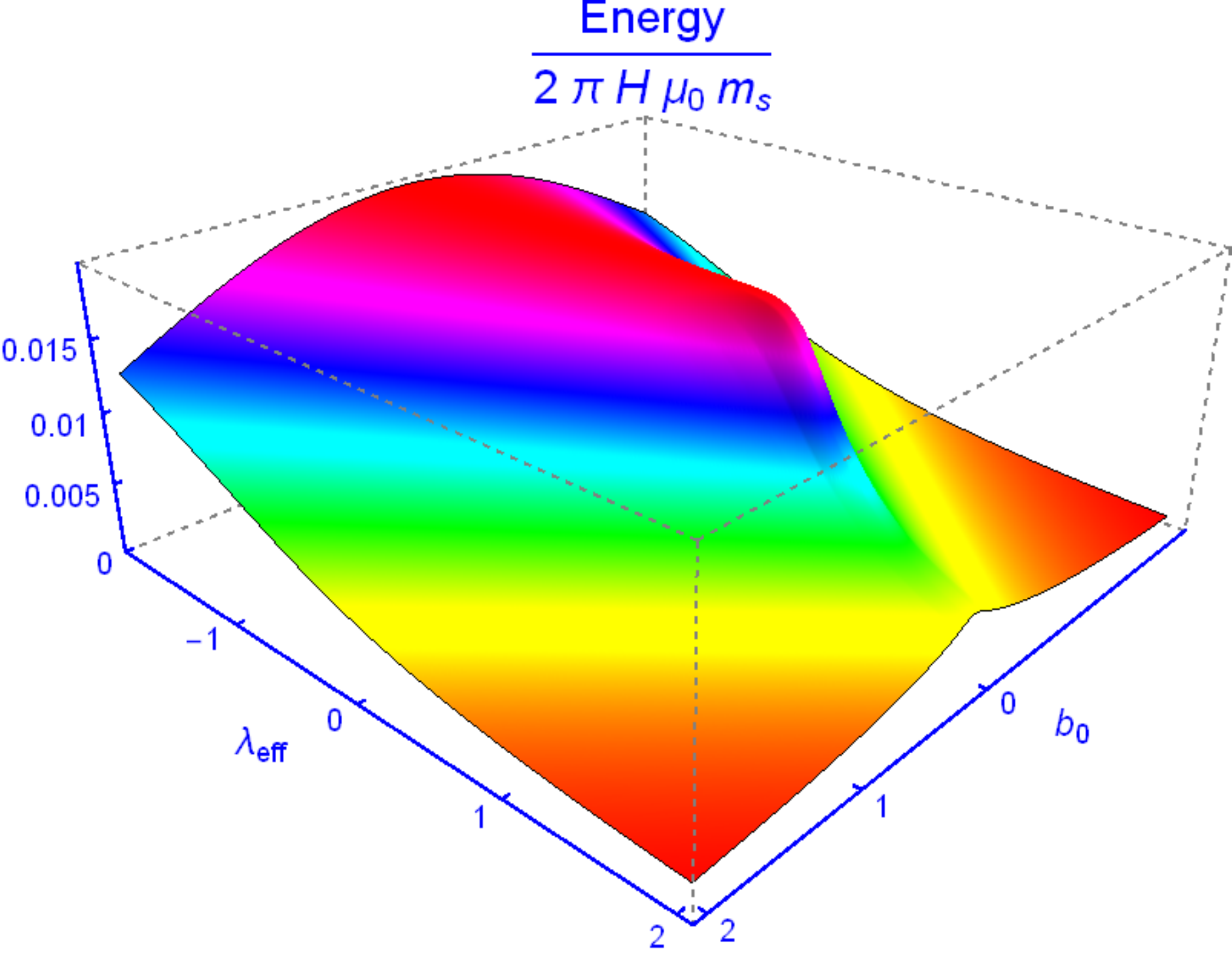}
\caption{(Color online). Dependence of the energy loss of the steady state solutions on $b_0$ and $\lambda_{\mr{eff}}$ for $\omega=0.01$, 
$\alpha_N = 0.1$ and $\omega_L = 0.2$.
\label{fig5}
} 
\end{center}
\end{figure}

\begin{figure}[ht] 
\begin{center} 
\includegraphics[width=7cm]{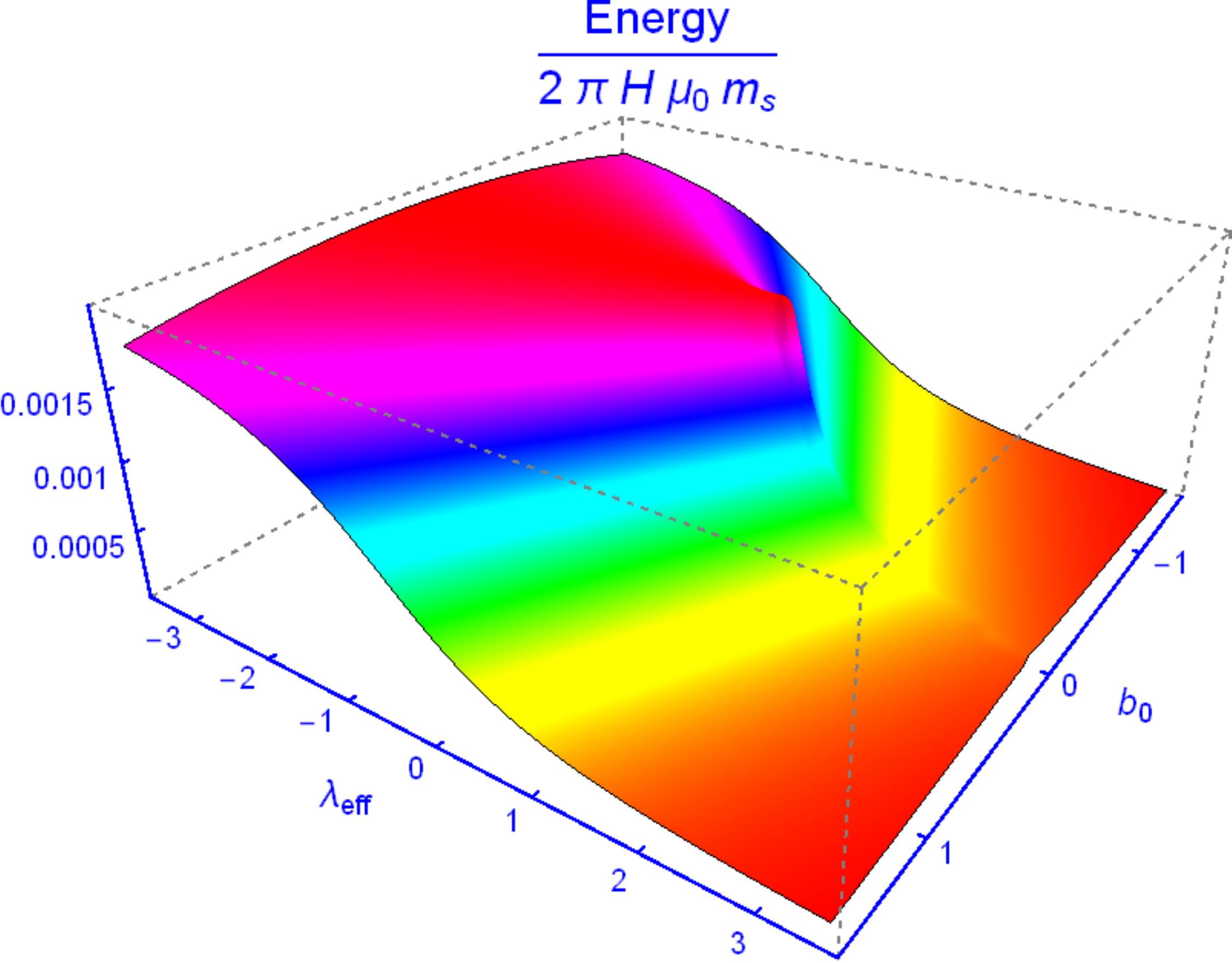}
\caption{(Color online). The same as \fig{fig5} but for $\omega=0.001$, i.e., the dependence of the energy loss of the steady state 
solutions on $b_0$ and $\lambda_{\mr{eff}}$.
\label{fig6}
} 
\end{center}
\end{figure}

\begin{figure}[ht] 
\begin{center} 
\includegraphics[width=10.5cm]{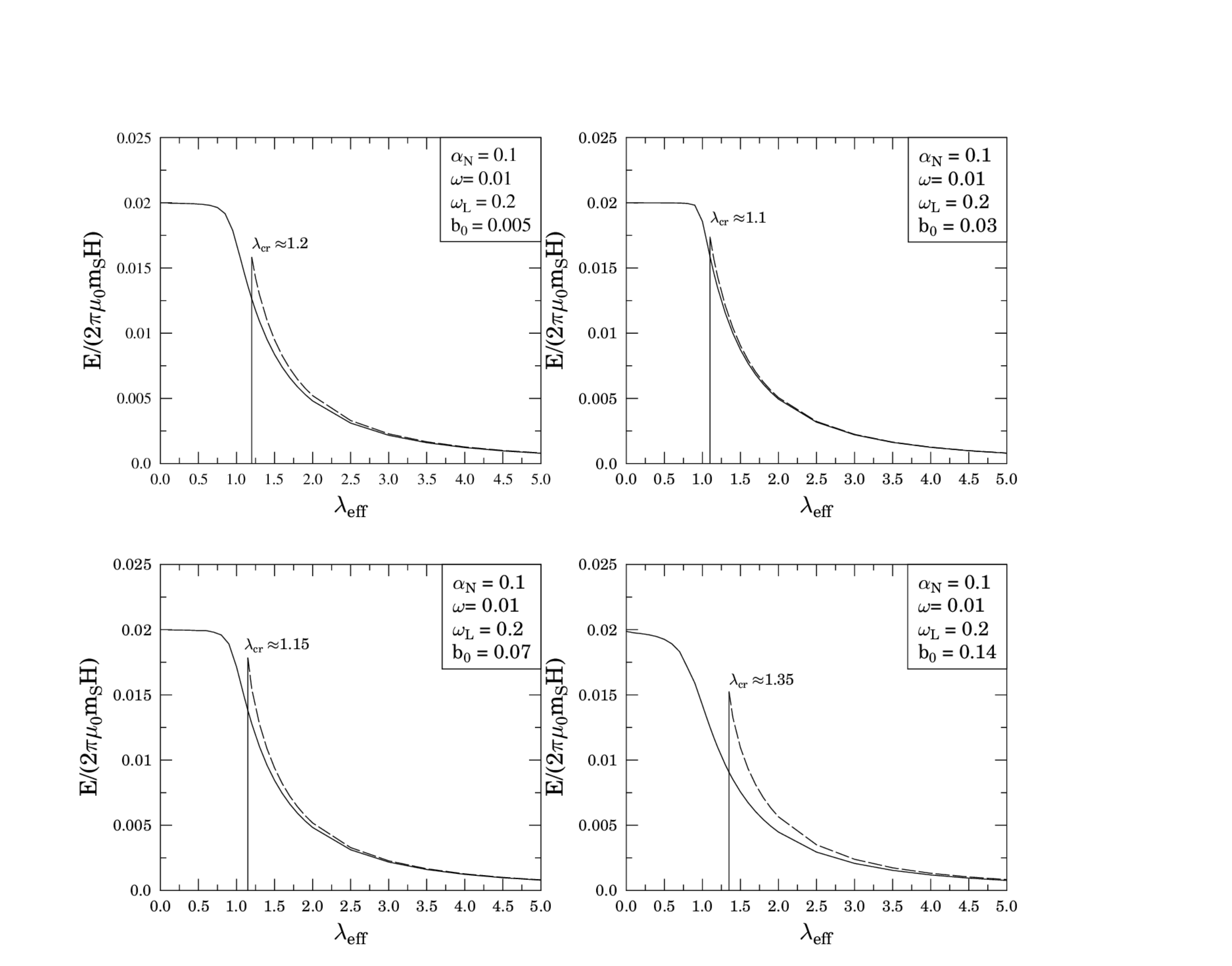}
\caption{Dependence of the energy loss of the steady state solutions on $b_0$ and $\lambda_{\mr{eff}}$.
\label{fig7}
} 
\end{center}
\end{figure}

Therefore, one expects a considerable dependence of the energy loss on $b_0$ which is demonstrated for relatively large \fig{fig5} 
%
%
and small \fig{fig6} frequencies.
%
%
In both cases one finds a decrease of the energy loss due to the presence of the static field independently its (positive or negative) 
direction which is in agreement with \eq{loss_low-freq_static}. In order to make this effect more visible, various "2D-slices" of the 3D 
plot for $\omega = 0.01$ is shown in \fig{fig7} which clearly demonstrates that any static field (perpendicular to the plane of rotation) 
decreases the energy loss obtained for steady state solutions.  
%
%
This unfortunate decreasing effect is even more recognisable and sharper near the critical value of the anisotropy which 
has serious drawback on the applicability of the enhancement mechanism described in the previous section since it requires 
special conditions, namely the anisotropy should be around its critical value.

\subsection{Out of the steady state}
\begin{figure}[ht] 
\begin{center} 
\includegraphics[width=7cm]{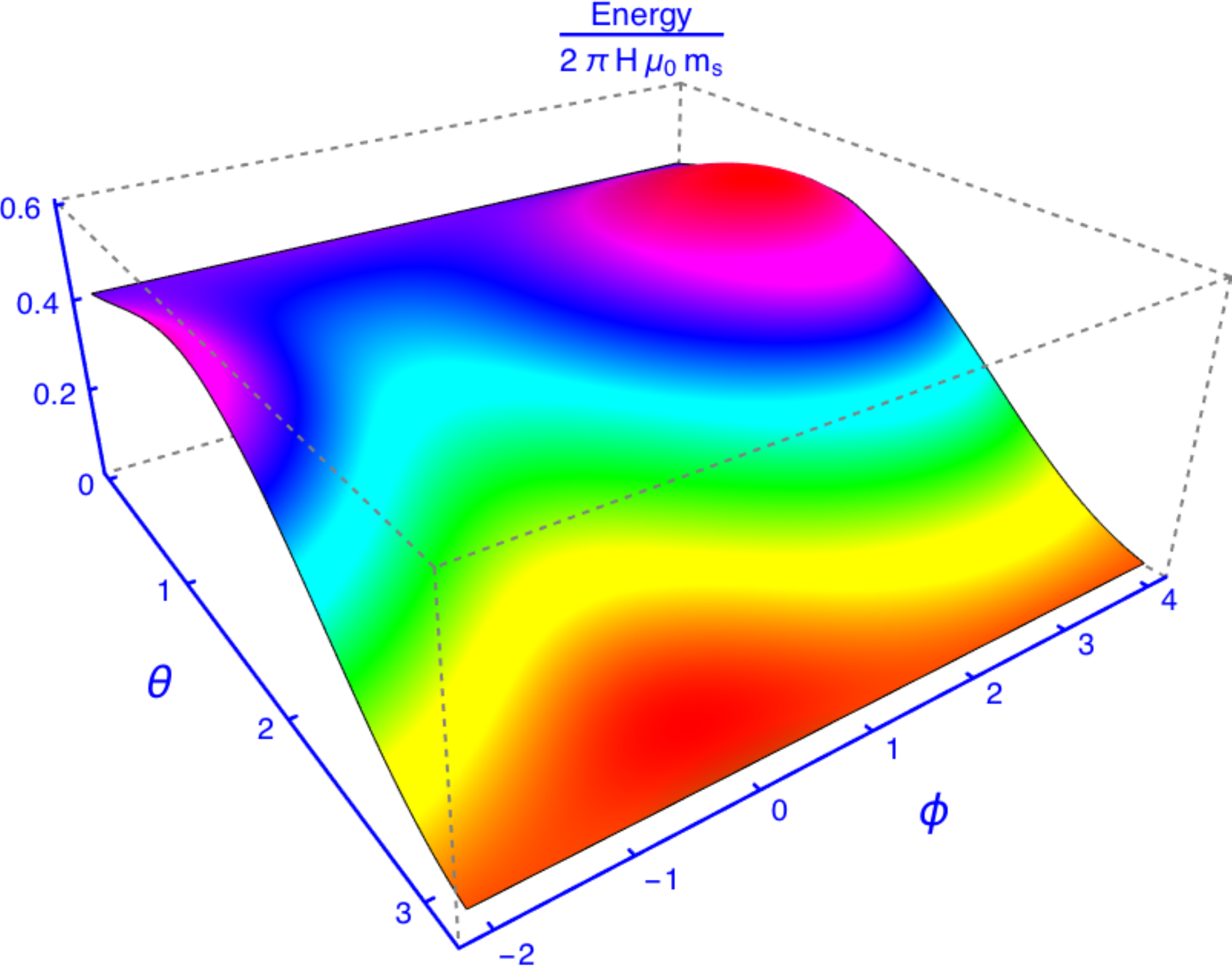}
\caption{(Color online). Same as \fig{fig2} but for $b_0 = -1.2$. 
\label{fig8}
} 
\end{center}
\end{figure}

Let us study the energy loss out of steady state.  The 3D-plot of \fig{fig8} shows the energy loss obtained in the first cycle of 
the external field as a function of various starting points on the ($\theta,\phi$) plane in the presence of a non-zero static field 
$b_0 \neq 0$. 
%
%
This signals that the static field influences the energy loss out of steady states significantly.

Furthermore, \fig{fig10} shows that even a relatively small static field modifies the energy loss obtained out of steady states. 
In particular, one finds a sudden jump in the (out of steady state) energy loss if it is calculated at the border of regions 
corresponds to different attractive fixed points. 
%
%
\begin{figure}[ht] 
\begin{center} 
\includegraphics[width=8.5cm]{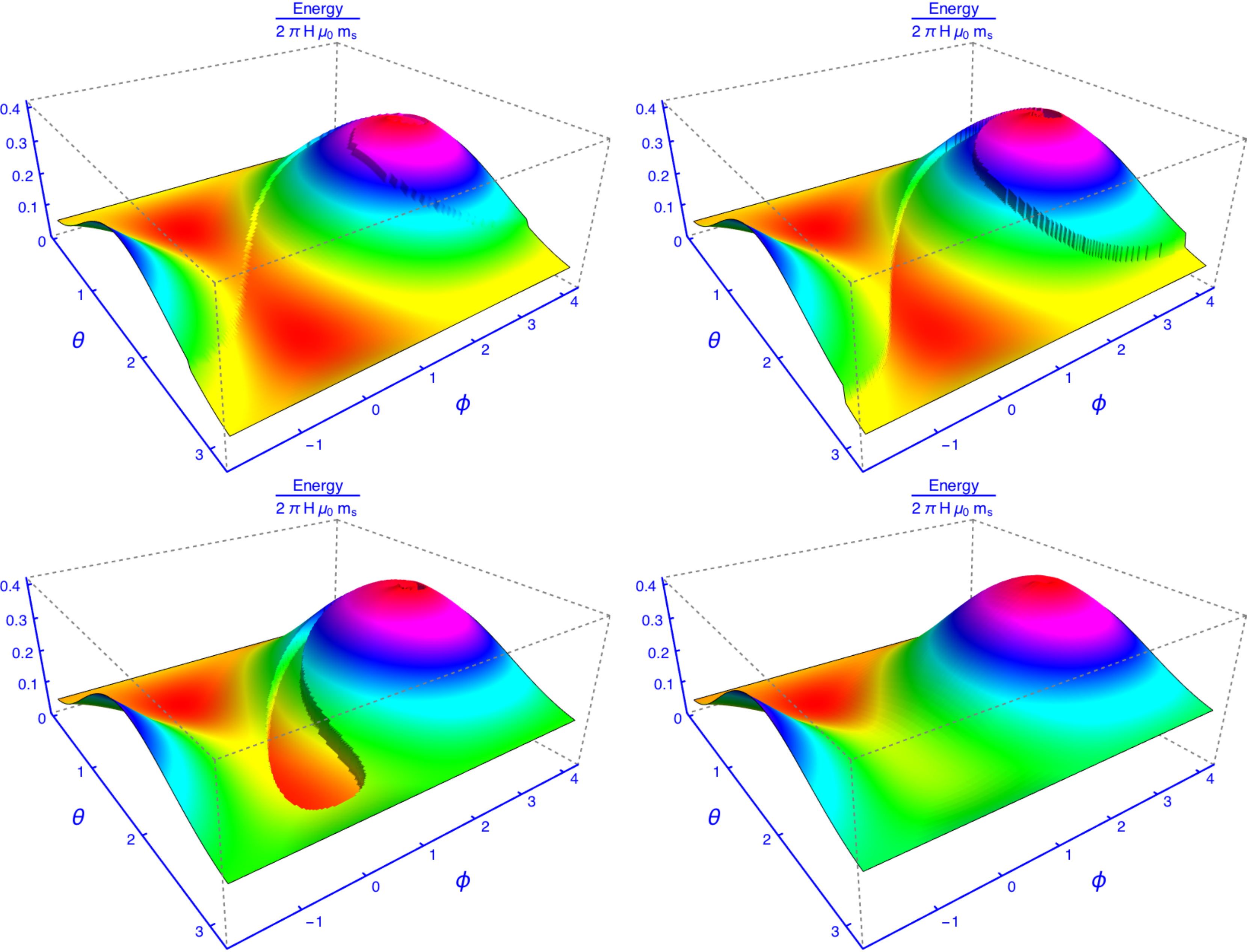}
\caption{(Color online). Sudden jumps in the energy loss appear along lines which separate regions corresponds to different 
attractive fixed points for various values of the static field $b_0 = 0.1, 0.2, 0.3, 0.4$ (from top left to bottom right) with 
$\alpha_N = 0.1, \omega_L = 0.2, \omega=0.01, \lambda_{\mr{eff}}= 1.7$.
\label{fig9}
} 
\end{center}
\end{figure}
In other words, a sudden jump of the energy loss can be seen along the lines which separates the attractive regions of the 
fixed points. It becomes more evident if one looks at various orbit maps in \fig{fig10} which correspond to various subgraphs 
of \fig{fig9}.  
%
%
\begin{figure}[ht] 
\begin{center} 
\includegraphics[width=8.5cm]{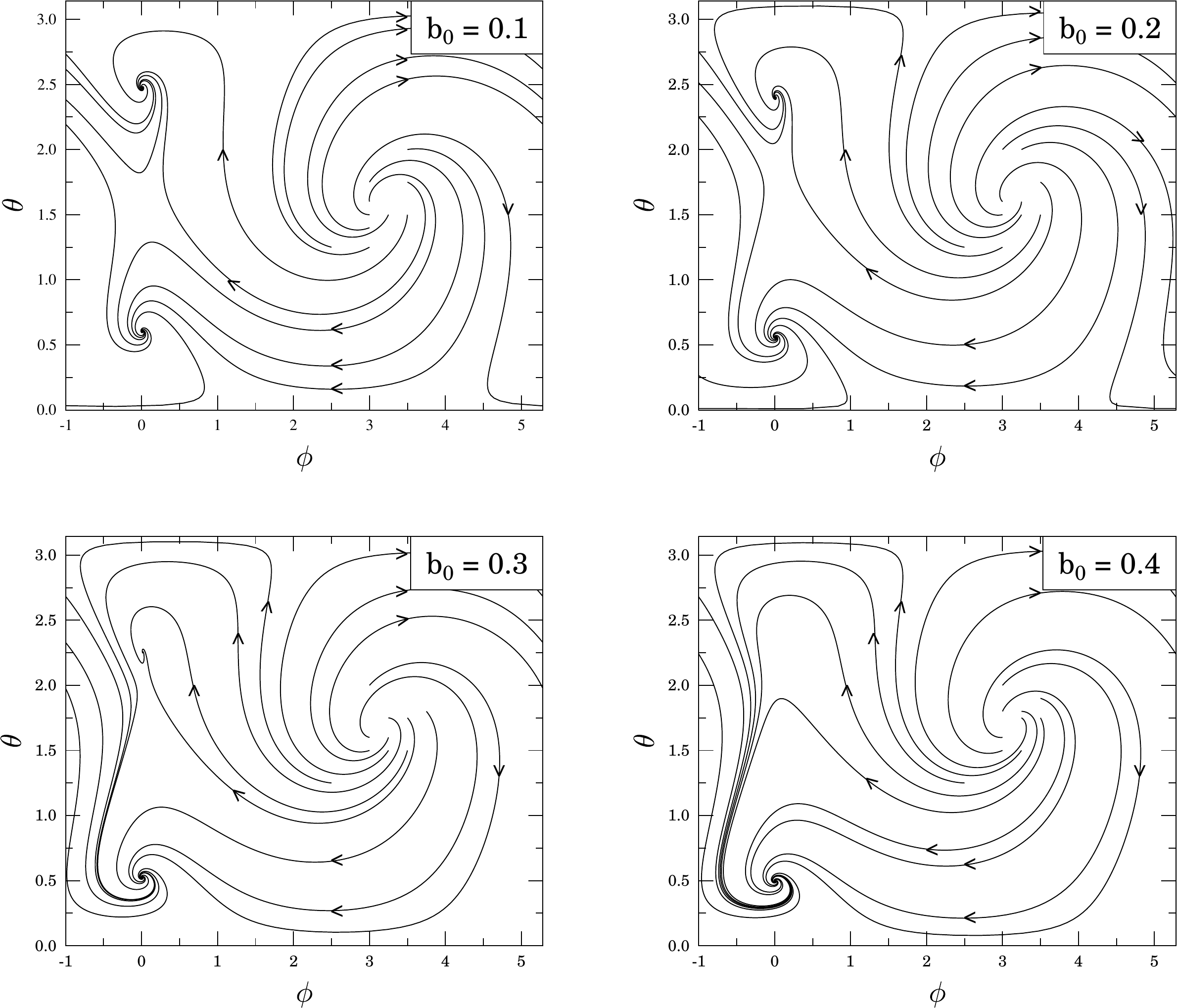}
\caption{Orbit maps correspond to subgraphs of \fig{fig9}.
\label{fig10}
} 
\end{center}
\end{figure}

\begin{figure}[ht] 
\begin{center} 
\includegraphics[width=7cm]{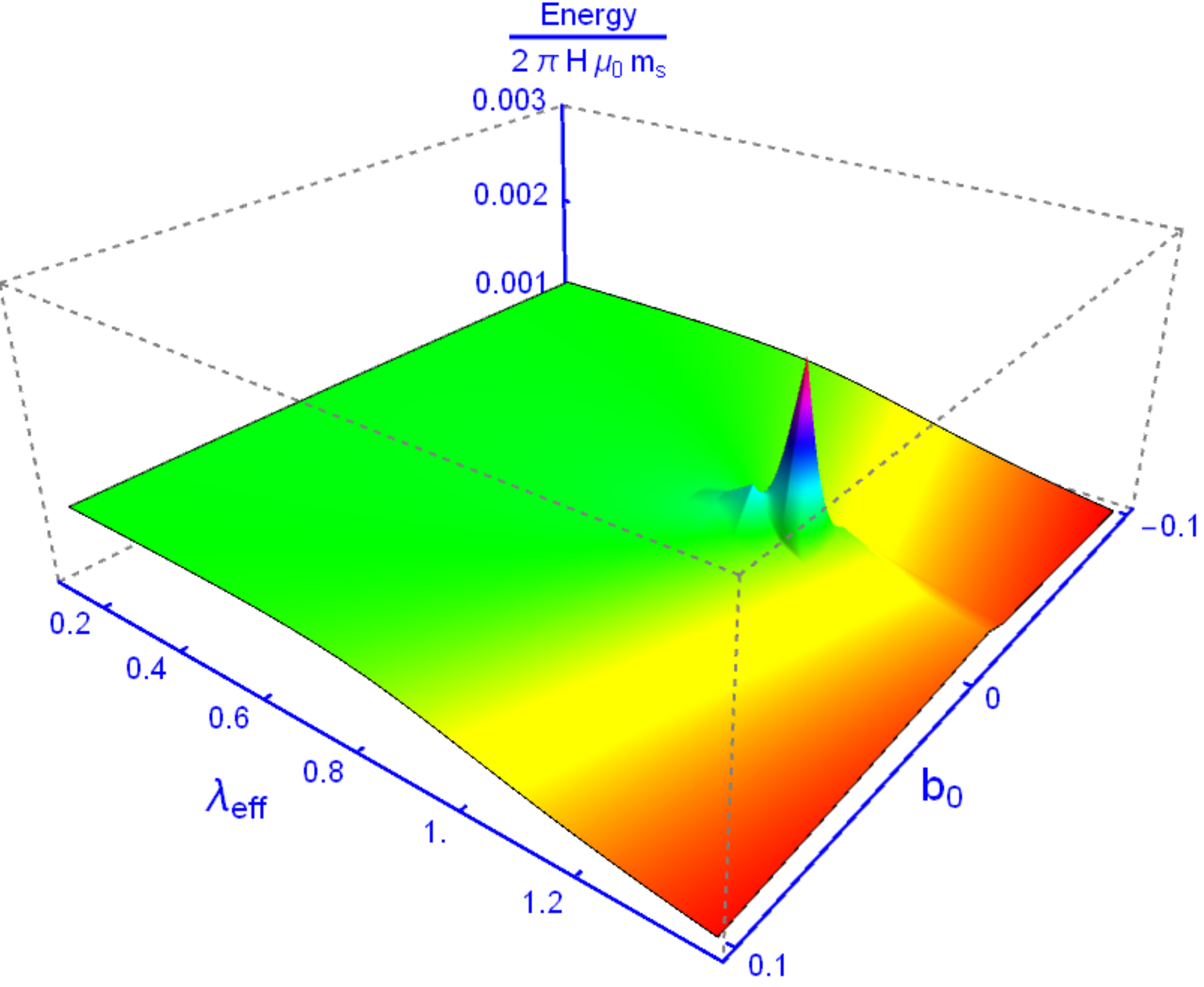}
\caption{(Color online). Dependence of the energy loss on $b_0$ and $\lambda_{\mr{eff}}$ for a rotating field with alternating 
direction for  $\alpha_N = 0.05, \omega_L = 0.2, \omega=0.001$.
\label{fig11}
} 
\end{center}
\end{figure}

As a final step we consider the dependence of the energy loss on $b_0$ obtained for the case of a rotating field with alternating 
direction where in every full cycle the angular velocity in \eq{H} changes its sign ($\omega \to -\omega$). Results are shown
in \fig{fig11}.
%
%
The sharp peak of \fig{fig11} around the critical value of the effective anisotropy demonstrate that the enhancement effect
strongly depends on the applied stabilising static field. In particular, for relatively small $b_0$ it vanishes. This has strong
constraint on the experimental realisation of the possible heat enhancement of the rotating field with alternating direction.
Requirements (e.g. particle size) which make possible to keep the orientation of the anisotropic nanoparticles for long 
enough time in the absence of any static field are discussed in \cite{heat_enhance}.

The findings discussed in this section are based on the approach of fixed nanoparticle which is an approximation where neither 
the thermal fluctuations, nor the motion of the particle has been taken into account and the direction of the static field is also fixed. 
Thus, an escape from the negative results of this section could be (i) the case when the direction of the static field is chosen 
arbitrarily, similarly to a randomly oriented anisotropy field \cite{Denisov_random}, (ii) the inclusion of thermal effects which allows 
the magnetic moment to jump between the two steady state solutions \cite{Denisov2006,Denisov_thermal}, (iii) the description 
when the motion of the particle as a whole \cite{lyutyy_general} is taken into account.

To complete the study of the static field, in the next section we try the first case, i.e., when the static field is assumed to be
in the plane of rotation.

\section{Static field in the plane of rotation}
\label{sec_nonzero_stat_in_plane}
In this section our goal is to clarify the role of a static field when the effective applied field is defined by 
\bea
\label{H_new}
{\bf H}_{\rm{eff}} = H \, \, \Big(\cos(\omega t) + b_0 + \lambda_{\mr{eff}} M_x,\,\,  \sin(\omega t), \,\, 0\Big), \hskip 1cm
H \equiv H_0  \hskip 0.5cm  {\mr{or}}   \hskip 0.5cm H \equiv \frac{H_0}{1+\vert b_0 \vert}
\eea
which means that the applied static and rotating fields and also the anisotropy field are in the XY plane. Important 
to note that the pre-factor $H$ is chosen either to be equal to $H_0$ or to $H_0/(1+\vert b_0\vert)$ in order
to keep the maximum of the applied field identical to the pure rotating one for the isotropic case ($\lambda_{\mr{eff}}$ =0).

\subsection{Isotropic case}
\begin{figure}[ht] 
\begin{center} 
\includegraphics[width=7cm]{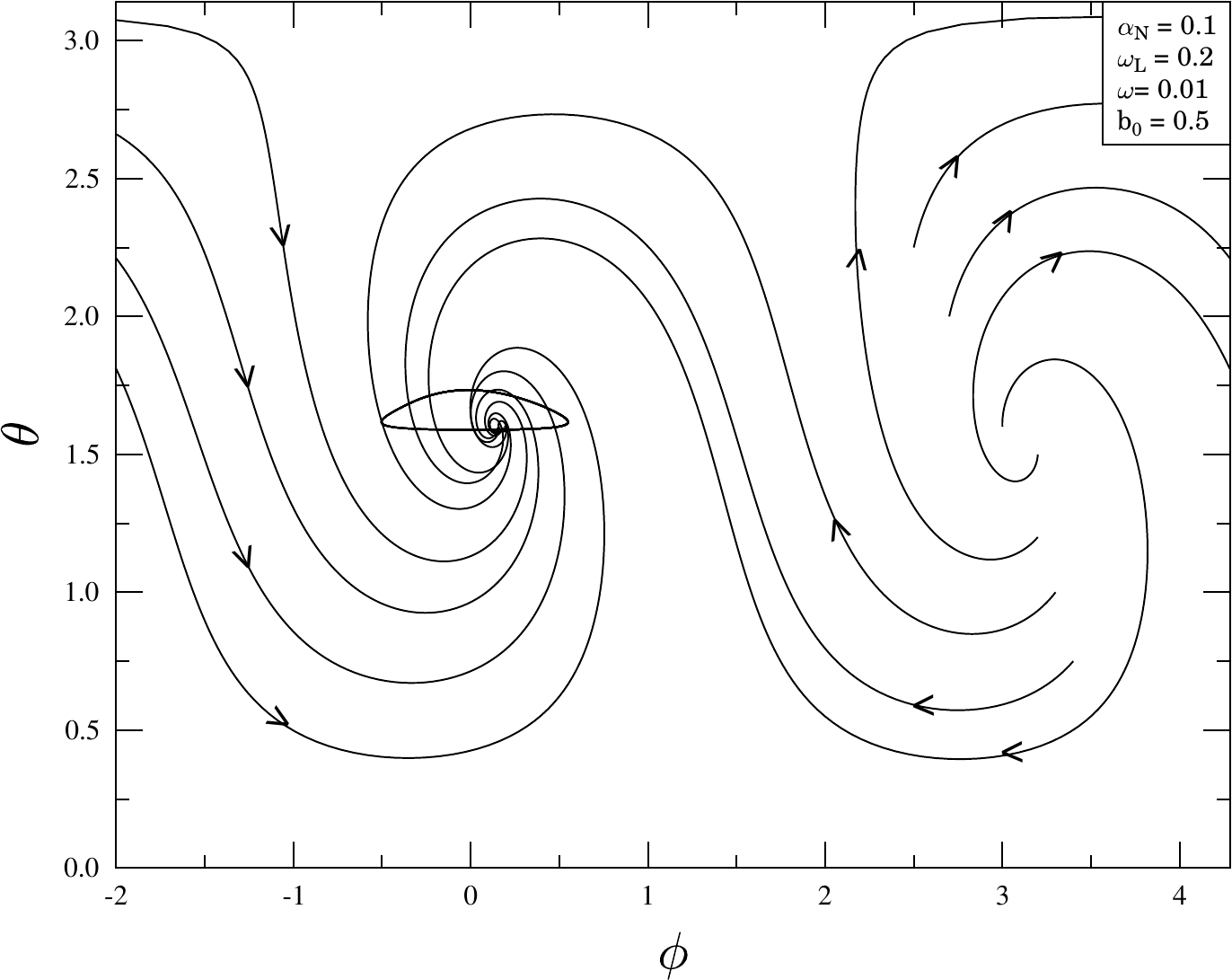}
\caption{Limit cycle (with a "triangle-like" shape) of the orbit map in the rotating frame for isotropic nanoparticles
($\lambda_{\mr{eff}}$ =0) when static field is assumed to be in the plane of rotation.
\label{fig12}
} 
\end{center}
\end{figure}

Let us immediately turn to the discussion of the isotropic case ($\lambda_{\mr{eff}}$ =0). The main difference in orbit maps 
(in the rotating frame) of the present and previous sections is that no fixed point solutions can be identified if the static field 
is {\em in} the plane of rotation. Instead, one finds attractive {\em limit cycles}, see \fig{fig12}.
%
%
Since the set of differential equations stands for the equation of motion of the magnetic moment has an explicit 
time-dependence, trajectories can cross each other. 

\begin{figure}[ht] 
\begin{center} 
\includegraphics[width=8.5cm]{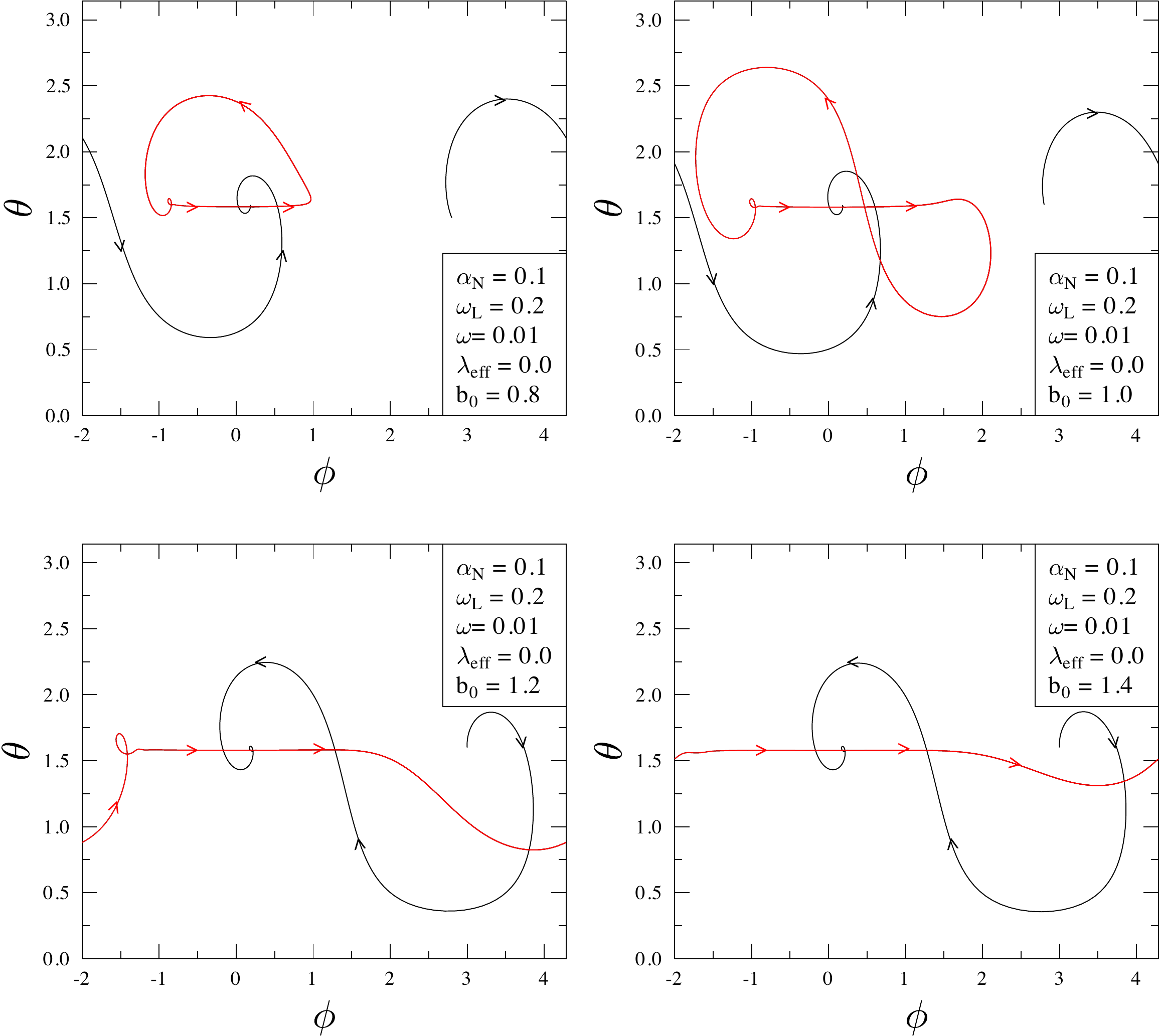}
\caption{(Color online). The shape of the limit cycle (red line) changes when the strength of the static field is modified.
\label{fig13}
} 
\end{center}
\end{figure}

\begin{figure}[ht] 
\begin{center} 
\includegraphics[width=12.5cm]{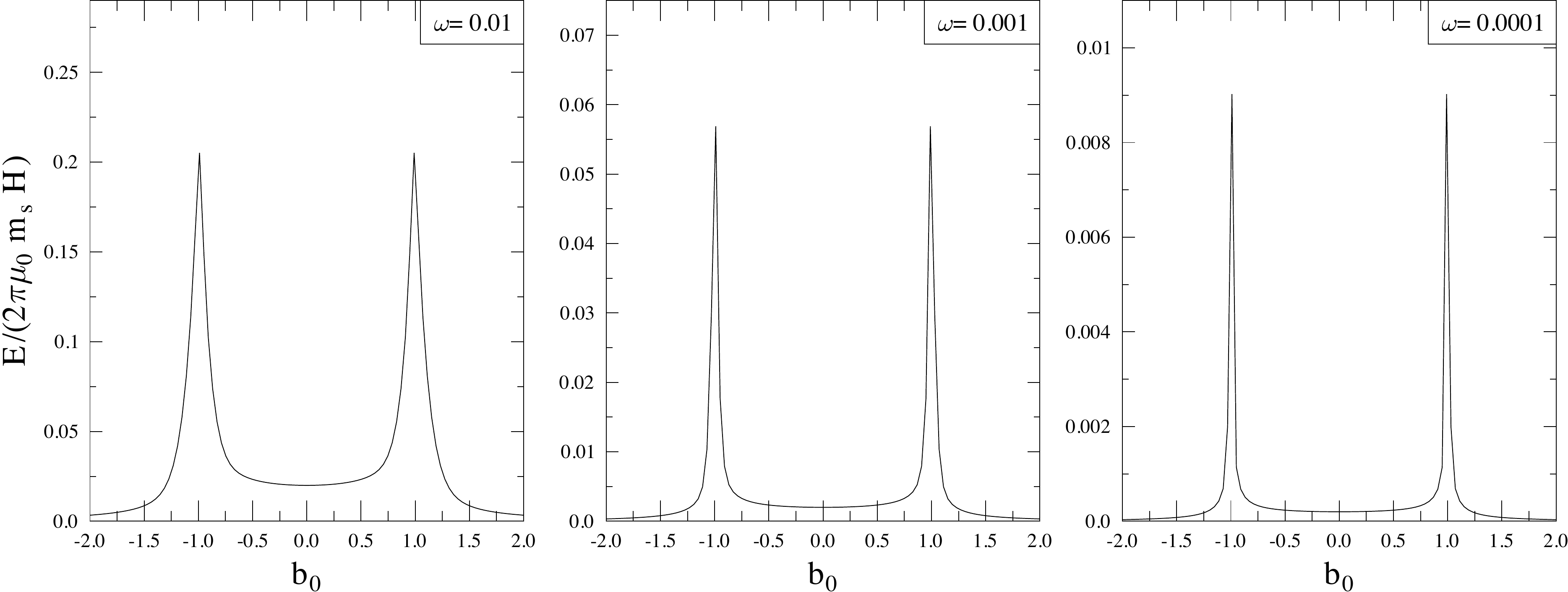}
\caption{Energy losses evaluated for various frequencies over limit cycles and plotted as a function of the amplitude 
$b_0$ of the static field which is in the plane of rotation. Parameters are chosen to be $\alpha_N = 0.1, \omega_L = 0.2, 
\lambda_{\mr{eff}}= 0$ and the energy is normalised by the factor $H \equiv H_0$.
\label{fig14}
} 
\end{center}
\end{figure}

The limit cycle depends on the strength of the static applied field. For example, \fig{fig13} shows how the shape of 
limit cycles depends on $b_0$.
%
%
The change in the shape is enhanced when $\vert b_0 \vert \sim 1$, i.e., when the amplitude of the static and the rotating fields
are comparable to each other. This motivates us to consider the energy loss/cycle obtained at the limit cycle around
the value $\vert b_0 \vert = 1$. Indeed, \fig{fig14} shows a large peak in the the energy loss at $\vert b_0 \vert =1$.
%
%
In the weak field limit ($\vert b_0 \vert \to 0$) the solution should recover the isotropic case. In the strong field limit 
($\vert b_0 \vert \to \infty$) one expect the vanish of the energy loss due to the extreme large static field which fixes the 
magnetic moment. The energy loss has a very large maximum in between the two limiting cases. 

\begin{figure}[ht] 
\begin{center} 
\includegraphics[width=12.5cm]{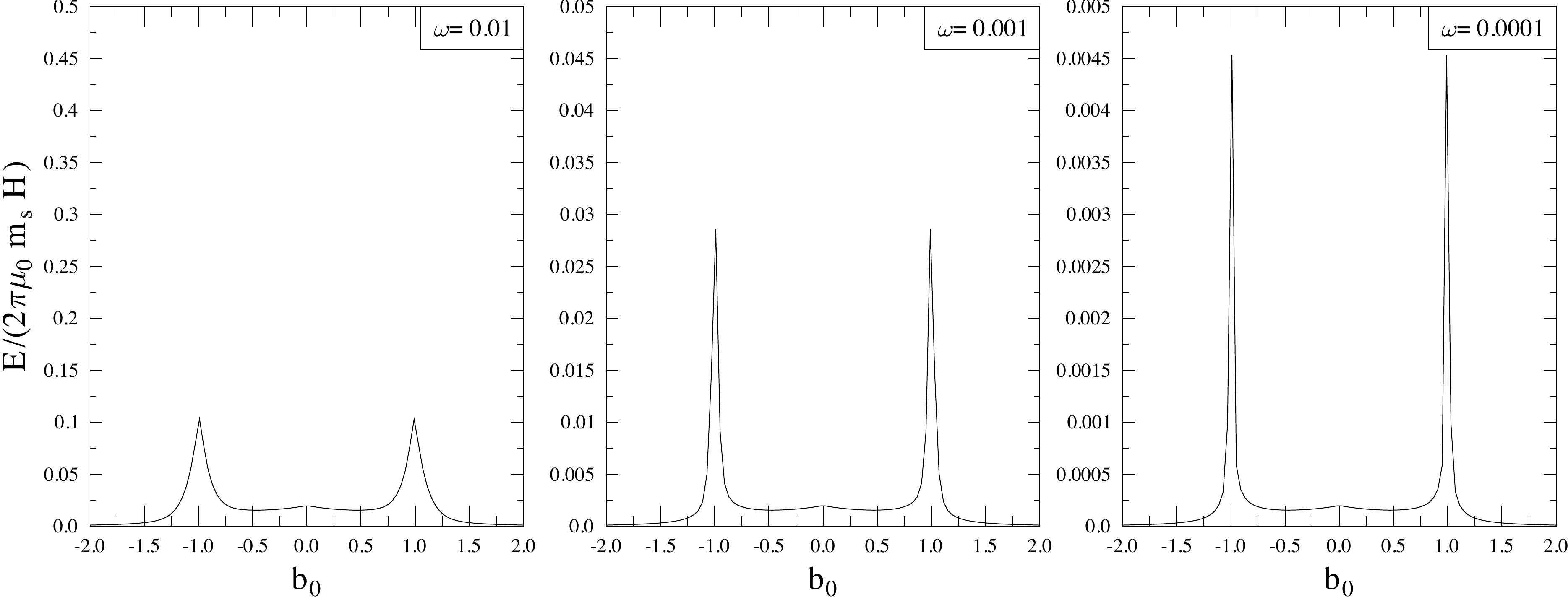}
\caption{The same as \fig{fig14} but the energy is normalised by the factor $H \equiv H_0/(1+\vert b_0\vert)$.
\label{fig15}
} 
\end{center}
\end{figure}

The same picture can be drawn if the the energy is normalised by the factor $H \equiv H_0/(1+\vert b_0\vert)$, see \fig{fig15}.
%
%

Moreover, the increase of the energy loss around $\vert b_0 \vert =1$ is enhanced in the limit of small frequency,
see the last figure of \fig{fig14}. In addition, the peak is sharpened. Therefore, in the limit of small frequencies 
required for hyperthermic treatments the enhancement effect reported in this work becomes larger and more 
sensible for the amplitude of the static field which can be used to "super-localise" the heat transfer: in case of an 
inhomogeneous applied static field, tissues are heated up only where the magnitudes of the static and rotating fields are 
equal to each other.

\subsection{Anisotropic case}
\begin{figure}[ht] 
\begin{center} 
\includegraphics[width=8.5cm]{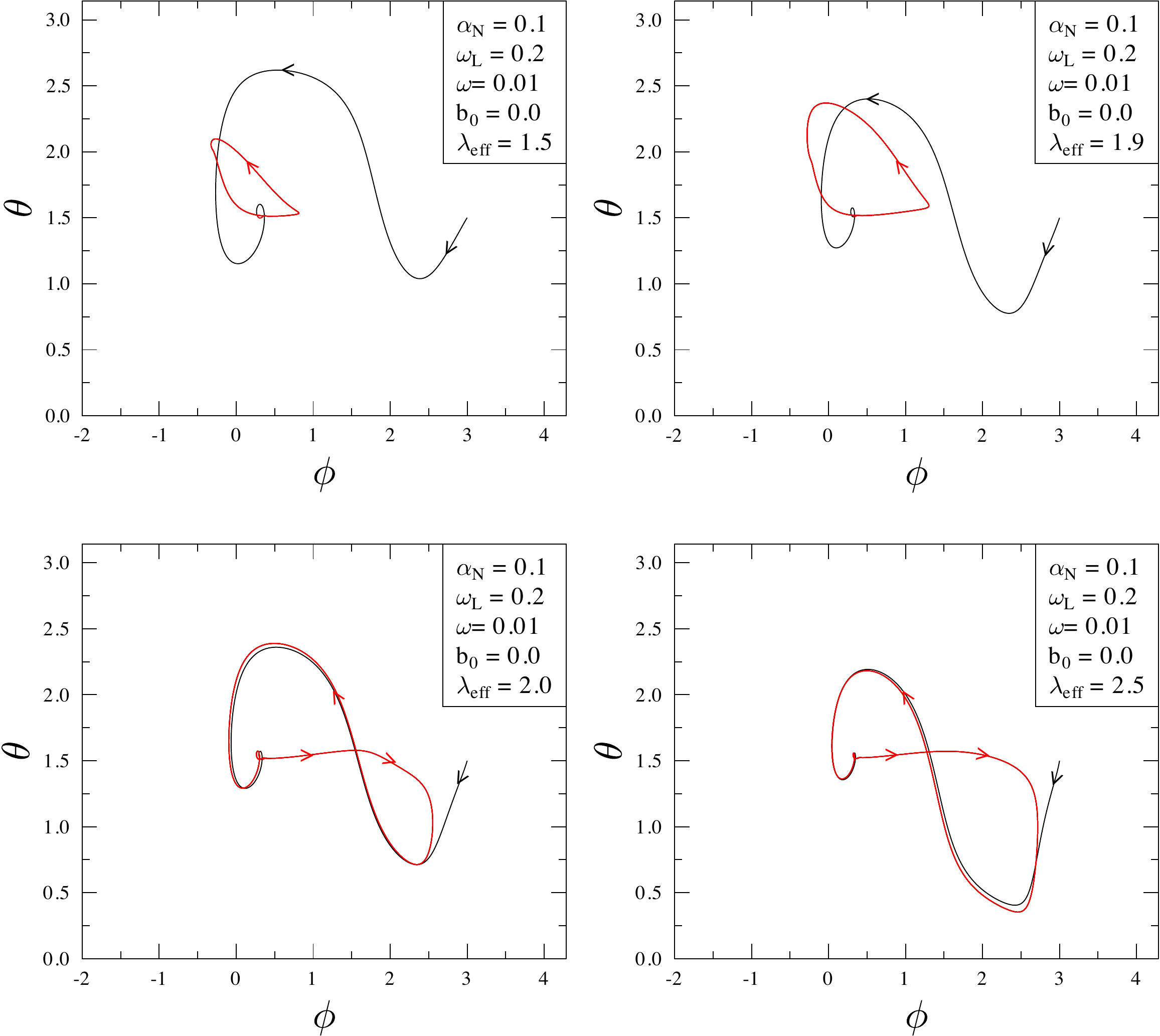}
\caption{(Color online). The shape of the limit cycle (red line) changes when $\lambda_{\mr{eff}}$ is modified (with $b_0 =0$).
\label{fig16}
} 
\end{center}
\end{figure}

Let us now consider the anisotropic case ($\lambda_{\mr{eff}} \neq 0$). The presence of anisotropy and the applied static field 
have a similar effect on orbit maps. One finds limit cycles which depend on the strength of the anisotropy parameter as shown 
by \fig{fig16} for $b_0 =0$.
%
%
A sudden change in the shape is observed when $\lambda_{\mr{eff}} \sim 2$. 

\begin{figure}[ht] 
\begin{center} 
\includegraphics[width=7.5cm]{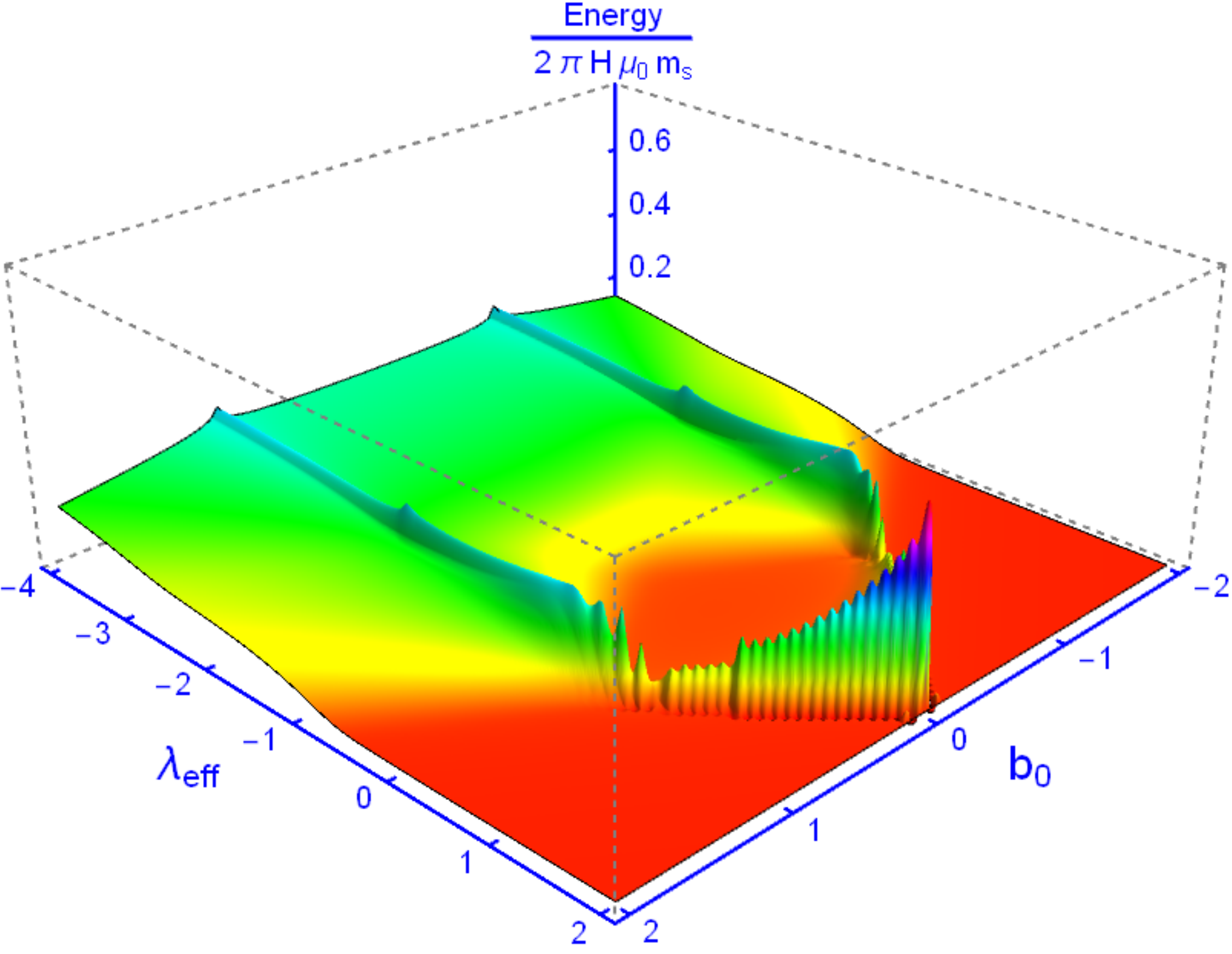}
\caption{(Color online). Energy losses evaluated (numerically) over a limit cycle and plotted as a function of the amplitude $b_0$ of the static field and 
the anisotropy parameter $\lambda_{\mr{eff}}$. Parameters are chosen to be $\alpha_N = 0.1, \omega_L = 0.2, \omega= 0.01$. 
Both the applied static and the anisotropy field is assumed to be in the plane of rotation. 
\label{fig17}
} 
\end{center}
\end{figure}

Due to symmetry reasons, the positive ($b_0>0$) and negative ($b_0<0$) static fields are expected to result in the same amount of 
energy losses. However, this is not true for the anisotropy, i.e., nanoparticles with negative ($\lambda_{\mr{eff}} < 0$) and 
with positive ($\lambda_{\mr{eff}} > 0$) anisotropy could have different behaviours. Indeed, the energy loss over the limit cycle depends 
on both $\lambda_{\mr{eff}}$ and $b_0$ shown by \fig{fig17} and there is a significant difference between the two cases.
%
%

When the anisotropy parameter is negative, i.e., considering nanoparticles with lens-shape geometric anisotropy, the energy loss
per cycle becomes large (compared to the case where $b_0 = 0$ and $\lambda_{\mr{eff}}=0$) and becomes more-or-less independent 
of the amplitude of the static field ($b_0$), specially for large anisotropy values where the energy loss per cycle saturates. Therefore, it 
seems to be a good choice, however, it is questionable whether the orientation of nanoparticles can be preserved in practice. 

For positive anisotropy the energy loss over the limit cycle has been enhanced only if $b_0$ and $\lambda_{\mr{eff}}$ fulfil the 
following relation
\beq
\vert b_0 \vert + \frac{1}{2} \lambda_{\mr{eff}} -1 =0.
\eeq
For the isotropic case ($\lambda_{\mr{eff}} =0$) it reduced to $\vert b_0 \vert -1 =0$ which indicates that the magnitudes of the static 
and rotating fields should be the same for enhancement ($b_0 = \pm 1$). This situation has been discussed in the previous subsection. 
For the case of vanishing static field ($b_0 =0$) the above relation gives $\frac{1}{2} \lambda_{\mr{eff}} -1 =0$ which is in agreement 
with the observation of a sudden change in the shape of limit cycles at $\lambda_{\mr{eff}} \sim 2$, see \fig{fig16}. According to \fig{fig17}, 
the largest enhancement effect is observed when $b_0 =0$ and $\lambda_{\mr{eff}}=2$, however, the orientation of the nanoparticles 
cannot be preserved, therefore, it has no relevance in practice. In general, the orientation matters for anisotropic nanoparticles, so, 
it is more reliable to use nanoparticles which are as isotropic as possible. 

Therefore, the best proposal for experimental realisation of the enhancement effect reported in this work is the use of {\em isotropic} 
nanoparticles where the enhancement of energy loss is expected for $b_0 = \pm 1$, i.e., when the magnitude of the static and rotating 
applied fields are the same. This is exactly the case discussed in the previous subsection.

Finally, let us note that new findings of the present work where enhancement effect is found are based on numerical solution of the
deterministic LLG equation. Analytical solutions are available in the literature based on perturbation theory, see \cite{Denisov_random},
but unfortunately it cannot be used to verify the enhancement effect of this work since it requires very large anisotropy (and also very
high frequencies), see \app{analytical}. However, our numerical results are verified by an analytical formula in the limit of vanishing 
static and anisotropy fields, $b_0 = \lambda_{\mr{eff}} = 0$, where the steady state solution \eq{steady_state} is used and we found
perfect agreement.

\section{Summary}
\label{sec_sum}
In this work we considered the effect of a static magnetic field on the energy losses obtained by magnetic nanoparticles
driven by a rotating applied field  in the framework of the deterministic Landau-Lifshitz-Gilbert equation. First, we studied a 
special orientation of nanoparticles where the anisotropy field (which characterises the uniaxial anisotropy) and also the 
applied static field are perpendicular to the plane of rotation. It was shown that energy losses related to steady state motions of 
magnetic moments are decreased significantly by a static applied field. 

Moreover, energy losses out of the steady states are also investigated in the presence of non-zero static fields using the above
orientation. This is motivated by the findings of Ref. \cite{heat_enhance} in which the heat generation is increased significantly 
by a rotating field where the direction of rotation changes periodically which pushes magnetic moment out of the steady state. 
In this work it was shown that the enhancement effect of \cite{heat_enhance} strongly depends on the applied static field. 
In particular, for relatively small $b_0$ it vanishes. This has strong constraint on the experimental realisation of the possible 
heat enhancement of the rotating field with alternating direction.

Based on the findings discussed above one can conclude that static field is always unfavourable, however, this is not the case. 
We showed that a significant increase in the energy loss/cycle is observed if the static field is situated {\em in} the plane of 
rotation when its strength is the same as the rotating one ($b_0 =1$). This enhancement is found to be much larger than that
of reported in \cite{heat_enhance} for the case of a rotating applied field with alternating direction. Therefore, the efficiency of
magnetic hyperthermia can be increased drastically by an applied static field combined with a rotating one if both are in the 
same plane and their magnitude are equal. This result is obtained for fixed nanoparticles when only the magnetic moment is 
allowed to move and no thermal effects are taken into account. In order to go beyond the approximations used
in this work one has to perform a more general study where the mechanical rotation of the particle and also the effect of thermal 
fluctuations are included (simultaneously) which is not discussed here since we consider the main result of the present work 
suitable for experimental tests (well separated nanoparticles in an aerogel matrix can serve as a good experimental setup).
Furthermore, the enhancement effect reported here depends on the magnitudes of the static and rotating fields very strongly 
(they should be equal otherwise the heating efficiency decreases rapidly). Therefore, an inhomogeneous static field can have 
a super-localising heating effect because it heats up tissues only where its magnitude is equal to the rotating one.

\section*{Acknowledgement}
This work is supported by the J\'anos Bolyai Research Scholarship of the Hungarian Academy of Sciences, by the European 
COST action TD1402 (RADIOMAG). The authors gratefully thank G. Gori and A. Trombettoni for useful discussions.

\appendix
\section{Perturbative analytical results}
\label{analytical}
The enhancement effect reported in this work requires a presence of a static and/or anisotropy fields in plane of rotation where 
the amplitudes of these fields satisfy a relation, i.e., $\vert b_0 \vert + \hf \lambda_{\mr{eff}} -1 = 0$ in case of positive anisotropy 
($\lambda_{\mr{eff}} >0$) where $b_0$ and $\lambda_{\mr{eff}}$ measure the strength of the static and the anisotropy fields to the 
rotating one. We argued that for technical reasons the isotropic case ($\lambda_{\mr{eff}} =0$) is more reliable for practice, but in 
principle an enhancement is found also for the anisotropic case even for vanishing static fields ($b_0 = 0$). Since our findings (even 
the above analytical formula) are based on numerical results it is useful to follow Ref. \cite{Denisov_random} where an analytical 
solution of the LLG equation in case of uniaxial anisotropy is presented in the framework of perturbation theory. The goal of this 
appendix is to show whether the method of Ref. \cite{Denisov_random} can be applied here to varify our numerical results.

In our case $\rho =1$ has to be chosen in Eq.~(2.1) and $\theta_a = \pi/2$, $\phi_a = 0$ in Eq.~(2.2) of \cite{Denisov_random}
which implies ${\bf e}_a = {\bf e}_x$ which fixes the easy axis for anisotropy. The perturbative treatment is based requiring a small 
value for the applied field amplitude compared to the anisotropy field, i.e., $h = H/H_a \ll 1$ which means $\lambda_{\mr{eff}} \gg 1$ 
in our notation. The solution ${\bf m}(t)$ of the LLG equation can be expressed as a series expansion,
\beq
{\bf m}(t) = \sum_{n=0}^\infty {\bf m}_n, \hskip 1cm \vert {\bf m}_n \vert \sim h^n,  \hskip 1cm {\bf m}_0 = {\bf e}_a = {\bf e}_x.
\eeq
One has to introduce a right-handed Cartesian coordinate system $x',y',z'$ characterised by the unit vectors 
${\bf e}_1, {\bf e}_2, {\bf e}_a$ which, in our case, ${\bf e}_y, {\bf e}_z, {\bf e}_x$. Then, the next-to-leading order solution to the LLG 
equation based on Eqs.~(3.29) and (3.26) of \cite{Denisov_random} by using $\kappa_a = \chi_a = 1$ and $\delta_a = \lambda_a = 0$ 
reads as
\beq
{\bf m}_1 =  \frac{h}{\Delta_{11}} \left[2\alpha \Omega^2  \sin{(\omega t)} + \Omega(1-l \Omega^2) \cos{(\omega t)} \right] {\bf e}_y
+ \frac{h}{\Delta_{11}} \left[(1-2\Omega^2 + l \Omega^2) \sin{(\omega t)} - \alpha \Omega(1+l \Omega^2) \cos{(\omega t)} \right] {\bf e}_z,
\eeq
where $\Delta_{11} = (1-l \Omega^2)^2 + 4 \alpha^2 \Omega^2$ and $l = 1+\alpha^2$ with $\Omega = \omega/(\omega_L \lambda_{\mr{eff}})$.
There is a further condition for the damping constant which should be small, i.e., $\alpha \ll 1$. The next-to-leading order solution 
exhibits a resonant behaviour near $\Omega = 1$ which requires a very large frequency for the applied field compared to the 
Larmor-frequency, i.e., $\omega \gg \omega_L$. This latter condition is certainly out of the range of hyperthermia where just the 
opposite relation is required. In addition, the enhancement effect of the present work requires comparable amplitudes, e.g., 
$\lambda_{\mr{eff}} = 2$ in case of a vanishing static field and $\vert b_0 \vert = 1$ for isotropic case. Therefore, the analytical 
method presented in Ref.~\cite{Denisov_random} cannot be applied to verify our numerical results.

\end{document}